\documentclass{jfm}
\usepackage{color}
\usepackage{amstext}
\usepackage{amssymb}
\usepackage{graphicx}
\usepackage{amsmath}

\newcommand\const{\mathrm{const}}

\newcommand\vF{\boldsymbol{F}}

\newcommand\vf{\boldsymbol{f}}

\newcommand\vn{\boldsymbol{n}}

\newcommand\vx{\boldsymbol{x}}

\newcommand\vr{\boldsymbol{r}}

\newcommand\vq{\boldsymbol{q}}
\newcommand\vQ{\boldsymbol{Q}}

\begin{document}

{\title[Self-propulsion of V-shape micro-robot] {Self-propulsion of V-shape micro-robot }}

\author[V. A. Vladimirov]
{V.\ns A.\ns V\ls l\ls a\ls d\ls i\ls m\ls i\ls r\ls o\ls v}

\affiliation{Dept of Mathematics, University of York, Heslington, York, YO10 5DD, UK}

\pubyear{2010} \volume{xx} \pagerange{xx-xx}
\date{Sept 12th 2012}

\setcounter{page}{1}\maketitle \thispagestyle{empty}

\begin{abstract}

In this paper we study the self-propulsion of a symmetric $V$-shape micro-robot (or $V$-robot) which consists
of three spheres connected by two arms with an angle between them; the arms' lengths and the angle are
changing periodically. Using an asymptotic procedure containing two-timing method and a distinguished limit,
we obtain analytic expressions for the self-propulsion velocity and Lighthill's efficiency. The calculations
show that a version of $V$-robot, aligned perpendicularly to the direction of self-swimming, is both the
fastest one and the most efficient one. We have also shown that such $V$-robot is faster and more efficient
than a linear three-sphere micro-robot. At the same time the maximal self-propulsion velocity of $V$-robots
is significantly smaller than that of comparable microorganisms.

\end{abstract}

\section{Introduction and formulation of problem\label{sect01}}

The studies of micro-robots represent a flourishing modern research topic which strives to create a
fundamental base for modern applications in medicine and technology, see \emph{e.g.}
\cite{Purcell, Koelher, NG+, Dreyfus, Yeomans1, Paunov, Lefebvre,  Gilbert, Golestanian, Golestanian1, Yeomans,
Pietro, Lauga}. The simplicity of both time-dependence and geometry of micro-robots represents their major
advantage in contrast with extreme complexity of self-swimming microorganisms,
\emph{e.g.} \cite{PedKes, VladPedl, Pedley, Polin}. This advantage allows to describe the motion of micro-robots
in a greater depth.
\begin{figure}
\centering\includegraphics[scale=0.9]{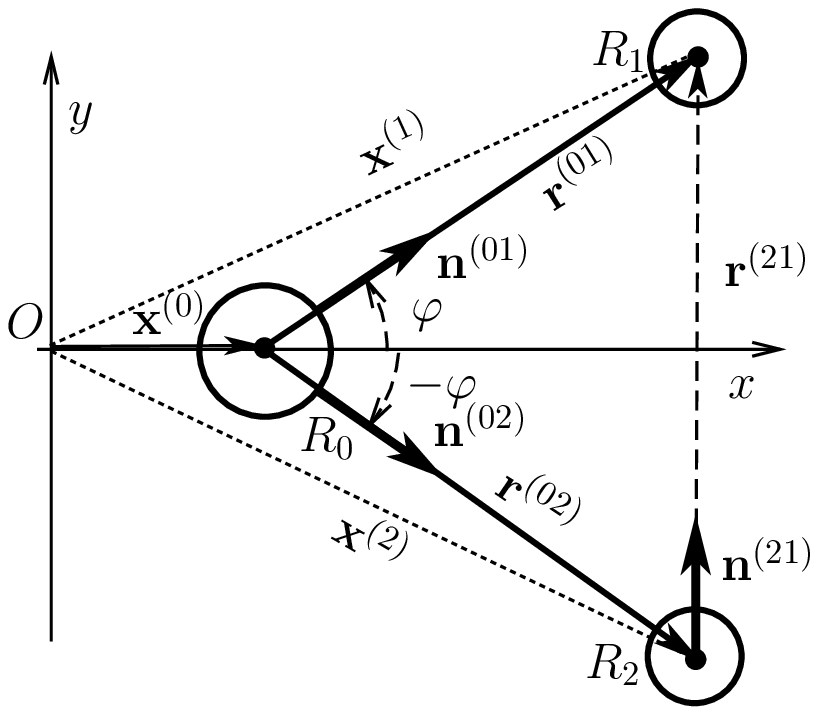}
\label{V-robot}
\end{figure}
In this paper, we study the self-propulsion of a symmetric $V$-shape three-sphere micro-robot (for brevity we
call it $V$-robot), see the figure. We employ two-timing method and distinguished limit arguments, which lead
to a simple and rigorous analytical procedure.  The self-propulsion velocity and Lighthill's swimming
efficiency are calculated analytically. It appears that $V$-robot can swim in both directions of $x$-axis,
and $V$-robot, aligned perpendicularly to the direction of self-swimming, is the most efficient one. In
addition, such $V$-robot is faster and more efficient than a linear three-sphere micro-robot, the most
studied one. $V$-robot has already been  studied numerically by \cite{Yeomans1}, but never analytically. Our
approach is technically different from all previous methods employed in the studies of micro-robots (except
\cite{VladimirovX3}). The possibility to derive an explicit formula (\ref{V}) for a $V$-shape three-sphere
micro-robot shows its strength and analytical simplicity. The used version of the two-timing method has been
developed in \cite{Vladimirov0,Vladimirov1,VladimirovMHD}.

We consider a symmetric $V$-shape  micro-robot ($V$-robot) consisting of three rigid spheres of radii
$R_\nu$, $\nu=0,1,2$ ($R_1=R_2$) connected by two arms of equal length $l$. The angle between the arms is
$2\varphi$. In plane cartesian coordinates the centers of the spheres and the distances between them are
$$
\vx^{(\nu)}=(x_1^{(\nu)},x_2^{(\nu)})\equiv(x^{(\nu)},y^{(\nu)}),\quad \vr^{(\mu\nu)}\equiv \vx^{(\nu)}-\vx^{(\mu)}
$$
We accept the notations (see the figure):
\begin{eqnarray}
&&\vx^{(0)}=(x^{(0)},y^{(0)})=(X,0),\quad \vx^{(1)}=(x^{(1)},y^{(1)})=(x,y)=(X+\xi,y),\label{notat-1}\\
&&\vx^{(2)}=(x^{(2)},y^{(2)})=(x,-y)=(X+\xi,-y); \nonumber\\
&&\vr^{(01)}\equiv (\xi,y),\quad \vr^{(02)}=(\xi,-y),
\quad \vr^{(21)}=(2y,0);\nonumber \label{notat-2}\\
&&|\vr^{(01)}|=|\vr^{(02)}|=l,\quad |\vr^{(21)}|=2y=2l\sin\varphi,\quad
(\xi,y)=l(\cos\varphi,\sin\varphi);\nonumber\\
&&\vn^{(01)}=(\cos\varphi,\sin\varphi),\quad\vn^{(02)}=(\cos\varphi,-\sin\varphi),\quad
\vn^{(21)}=(0,1)\nonumber
\end{eqnarray}
where $\vn^{(\mu\nu)}$ are unit vectors along $\vr^{(\mu\nu)}$. We use latin subscripts ($i,k=1,2$) for
cartesian components of vectors and tensors, subscript $\alpha=1,2,3$ -- for generalised coordinates, and
subscripts (or superscripts) $\mu,\nu=0,1,2$ to identify the spheres. $V$-robot represents a mechanical
system described by three scalar parameters (generalised coordinates)
\begin{eqnarray}
&&\vq=(q_1,q_2,q_3)\equiv(X,l,\varphi)\label{q}
\end{eqnarray}
$V$-robot moves due to the prescribed motion of the arms
\begin{eqnarray}
&&l=L+\varepsilon\widetilde{l}(\tau), \quad \varphi=\Phi+\varepsilon\widetilde{\varphi}(\tau);\quad
\tau\equiv\omega t,\
\omega=\const,\quad \varepsilon=\const\label{constraints}
\end{eqnarray}
where the functions $\widetilde{l}(\tau)$ and $\widetilde{\varphi}(\tau)$ are $2\pi$-periodic with  zero
average values, while $L$ and $\Phi$ are constants. The spheres experience external friction forces
$\vF^{(\nu)}=(F_1^{(\nu)},F_2^{(\nu)})$; however, the arms are chosen to be so thin in comparison with any
$R_\nu$, that their interaction with a fluid is negligible.

The considered problem contains three characteristic lengths: the length of arms $L_{\text{char}}$, the
radius  of spheres $R_{\text{char}}$, and amplitude  of the arm's oscillations $a_{\text{char}}$. The
characteristic time-scale is $T_{\text{char}}\equiv 1/\omega$ and the characteristic force is
$F_{\text{char}}$. We have chosen
\begin{eqnarray}\label{scales}
&&L_{\text{char}}\equiv L,\quad R_{\text{char}}\equiv (R_0+2R_1)/3,\  a_{\text{char}}\equiv\varepsilon L,\
F_{\text{char}}\equiv 6\pi\eta R_{\text{char}}L_{\text{char}}/T_{\text{char}}
\end{eqnarray}
where $\eta$ is viscosity of a fluid. Two small parameters are
\begin{eqnarray}\label{small-par}
&&\varepsilon\equiv a_{\text{char}}/L\ll 1,\quad \delta\equiv 3R_{\text{char}}/(2L)\ll 1
\end{eqnarray}
The dimensionless (asteriated) variables  are chosen as
\begin{eqnarray}
&& \vx=L\vx^*,\quad R_i=R_{\text{char}} R_i^*,\quad  t=T_{\text{char}}t^*,\quad
 f_i=F_{\text{char}}f_i^*;
\label{scales1}
\end{eqnarray}
Below we use only dimensionless variables, however all asterisks are omitted.

Generalized coordinates $\vq=\vq(t)$ (\ref{q}) determine the motion of $V$-robot. It can be described by the
Lagrangian function $\mathcal{L}=\mathcal{L}(\vq,\vq_t)$, which includes the constraints (\ref{constraints})
\begin{eqnarray}
&&\mathcal{L}(\vq,\vq_t)=\mathcal{K}+f(l-1-\varepsilon\widetilde{l})+g(\varphi-\Phi-\varepsilon\widetilde{\varphi})
\label{Lagr-contstr}
\end{eqnarray}
where subscript $t$ stands for $d/dt$, $f$ and $g$ are Lagrangian multipliers, and $\mathcal{K}$ is kinetic
energy of a robot; notice that in dimensionless variables $L=1$. The Lagrange equations are
\begin{eqnarray}
&&\frac{d}{dt}\frac{\partial \mathcal{L}}{\partial q_{\alpha t}} -\frac{\partial \mathcal{L}}{\partial
q_\alpha}=Q_\alpha,\quad Q_\alpha=\sum_{\nu=0}^2\sum_{j=1}^2 F_j^{(\nu)}\frac{\partial x_j^{(\nu)}}{\partial
q_\alpha}\label{Lagr-eqns}
\end{eqnarray}
where $\vQ=(Q_1,Q_2,Q_3)$ is the generalized external force, exerted by a fluid on $V$-robot.

The fluid flow past $V$-robot is described by the Stokes equations. In consistent approximation masses of
spheres and arms are negligible, then $\mathcal{K}\equiv 0$. Hence (\ref{Lagr-contstr}),(\ref{Lagr-eqns})
produce the system of equations
\begin{eqnarray}
&&Q_1=0,\quad Q_2+f=0,\quad Q_3+g=0 \label{eqns-main-simple}
\end{eqnarray}
The calculations of $Q_\alpha$ (\ref{Lagr-eqns}) with the use of (\ref{notat-1}) yield
\begin{eqnarray}\label{interforces}
&&Q_1=F_1^{(0)}+F^+ ,\quad Q_2=F^+\cos\varphi + F^-\sin\varphi,\\
&&Q_3=-l F^+\sin\varphi+ l F^-\cos\varphi; \nonumber\\
&&F^+\equiv F_1^{(1)}+F_1^{(2)},  \quad F^-\equiv F_2^{(1)}-F_2^{(2)}\nonumber
\end{eqnarray}
The substitution of (\ref{interforces}) into (\ref{eqns-main-simple}) and simple transformations yield
\begin{eqnarray}
&&F_1^{(0)}+F^+=0,\quad F^+=\lambda,\quad F^-=\sigma; \label{eqns-main-final}\\
&&\lambda\equiv-f\cos\varphi+(g/l)\sin\varphi,\quad\sigma\equiv-(g/l)\cos\varphi-f\sin\varphi, \nonumber
\end{eqnarray}
The explicit expressions for $\vF^{(\nu)}$ are
\begin{eqnarray}\label{forces-Stokes}
&&\vF^{(\nu)}\simeq -R_\nu\vx_t^{(\nu)}+
\delta \sum_{\mu\neq\nu}R_\mu R_\nu\mathbb{S}^{(\nu\mu)}\vx_t^{(\mu)}\\
&&\mathbb{S}^{(\mu\nu)}=S_{ik}^{(\mu\nu)}\equiv\frac{1}{|\vr^{(\mu\nu)}|}(\delta_{ik}+n_i^{(\mu\nu)}n_k^{(\mu\nu)})\nonumber
\end{eqnarray}
Each force $\vF^{(\nu)}$ represents the first approximation for the Stokes friction force exerted on a sphere
moving in the flow generated by the other two spheres. To construct (\ref{forces-Stokes}) we use a classical
explicit formula for a fluid velocity past a moving sphere, see
\cite{Lamb,Landau,Moffatt}.
The substitution of (\ref{forces-Stokes}) into  (\ref{eqns-main-final}) yields:
\begin{eqnarray}
&&R_0x_{1t}^{(0)}+R_1(x_{1t}^{(1)}+x_{1t}^{(2)})-\delta A=0
\label{eqns-tobe-solved1}\\
&&R_1(x_{1t}^{(1)}+x_{1t}^{(2)})-\delta B=\lambda
\nonumber\label{eqns-tobe-solved2}\\
&&R_1(x_{2t}^{(1)}-x_{2t}^{(2)})-\delta C=\sigma
\nonumber\label{eqns-tobe-solved3}
\end{eqnarray}
where
\begin{eqnarray}
&&A\equiv R_0R_1\left[S_{1k}^{(01)}x_{kt}^{(1)}+S_{1k}^{(02)}x_{kt}^{(2)}+
(S_{1k}^{(01)}+S_{1k}^{(02)})x_{kt}^{(0)}\right] +R_1^2S_{1k}^{(21)}(x_{kt}^{(1)}+x_{kt}^{(2)})\nonumber\\
&&S_{1k}^{(01)}=\frac{1}{l}\left(1+{\xi^2}/{l^2}, {\xi y}/{l^2}\right),\
S_{1k}^{(02)}=\frac{1}{l}\left(1+{\xi^2}/{l^2}, -{\xi y}/{l^2}\right),\ S_{1k}^{(21)}=\frac{1}{2y}\left(0,
1\right)\nonumber\label{formula-A}
\end{eqnarray}
with summation convention over $k=1,2$. Functions  $B$ and $C$ will not affect the calculation of the
self-propulsion velocity with precision linear in $\delta$, therefore it is sufficient to write $B=O(1)$ and
$C=O(1)$. The equations (\ref{eqns-tobe-solved1}),(\ref{constraints}), (where definitions (\ref{notat-1})
should be taken into account) represent a system of five equations for five unknown functions of time:\ $X$,
$\xi$, $y$, $\lambda$, and $\sigma$.

\section{Two-timing method and asymptotic procedure \label{sect04}}

\subsection{Functions and notations}

The following \emph{dimensionless} notations and definitions are in use:

\noindent
(i) $s$ and $\tau$ denote slow time and fast time;  subscripts $\tau$ and $s$ stand for  related partial
derivatives.

\noindent
(ii) A dimensionless function, say $G=G(s,\tau)$, belongs to class $\cal{I}$ if $G={O}(1)$ and all  partial
$s$-, and $\tau$-derivatives of $G$ (required for our consideration) are also ${O}(1)$. In this paper all
functions belong to   class $\cal{I}$, while all small parameters appear as explicit multipliers.

\noindent
(iii) We consider only \emph{periodic in $\tau$ functions} $
\{G\in  \mathcal{P}:\quad G(s, \tau)=G(s,\tau+2\pi)\},
$ where $s$-dependence is not specified. Hence all considered below functions belong to $\cal{P}\bigcap
\cal{I}$.

\noindent
(iv) For  arbitrary $G\in \cal{P}$ the \emph{averaging operation} is
\begin{eqnarray}
\langle {G}\,\rangle \equiv \frac{1}{2\pi}\int_{\tau_0}^{\tau_0+2\pi}
G(s, \tau)\,d\tau\equiv \overline{G}(s),\qquad\forall\ \tau_0\label{oper-1}
\end{eqnarray}

\noindent
(v)  \emph{The tilde-function} (or purely oscillating function) represents a special case of
$\cal{P}$-function with zero average $\langle\widetilde G \,\rangle =0$. The \emph{bar-function} (or
mean-function) $\overline{G}=\overline{G}(s)$ does not depend on $\tau$. A unique decomposition
$G=\overline{G}+\widetilde{G}$ is valid.

\subsection{Asymptotic procedure and successive approximations}

The introduction of  fast time variable $\tau$ and slow time variable $s$ represents a crucial step in our
asymptotic procedure. We choose $\tau=t$ and $s=\varepsilon^2 t$. This choice can be justified by the same
distinguished limit arguments as in \cite{VladimirovMHD}. Here we present this choice without proof, however
the most important part of this proof (that this choice  leads to a valid asymptotic procedure) is exposed
and exploited below. We use the chain rule
\begin{eqnarray}\label{chain}
&&d/dt=\partial/\partial\tau+\varepsilon^2\partial/\partial s
\end{eqnarray}
and then we  accept (temporarily) that $\tau$ and $s$ represent two independent variables. Unknown functions
$X$, $\xi$, $y$, $\lambda$, and $\sigma$  are taken as regular series
\begin{eqnarray}\label{x-f-ser}
&&X(\tau,s)=X_0(\tau,s)+\varepsilon X_1(\tau,s)+\varepsilon^2 X_2(\tau,s)+\dots
\end{eqnarray}
and similar expressions for $\xi$, $y$, $\lambda$, and $\sigma$. In (\ref{x-f-ser}) we take
$\widetilde{X}_0\equiv 0$, which  which express the basic property of considered solutions: long distances of
self-swimming are caused by small oscillations. After the application of (\ref{chain}) to (\ref{x-f-ser}) we
have
\begin{eqnarray}\label{x-f-ser1}
&&X_t=\varepsilon \widetilde{X}_{1\tau}+\varepsilon^2
(\widetilde{X}_{2\tau}+\overline{X}_{0s})+O(\varepsilon^2)
\end{eqnarray}
and similar expressions for $\vx^{(\nu)}$.

The successive approximations of equations (\ref{eqns-tobe-solved1}) are:

\noindent \emph{Terms of order} $\varepsilon^0=1$:\  $\lambda_0\equiv 0$ and $\sigma_0\equiv 0$;

\noindent  \emph{Terms of order}  $\varepsilon^1$:
\begin{eqnarray}
&&{3}\widetilde{X}_{1\tau}=-2R_1\widetilde{\xi}_{1\tau},
\quad \widetilde{X}_{1\tau}+\widetilde{\xi}_{1\tau}=\lambda_{1}/2,\quad \widetilde{y}_{1\tau}=
\sigma_{1}/2\label{first-order}
\end{eqnarray}
Its average part gives $\overline{\lambda}_1\equiv 0$ and $\overline{\sigma}_1\equiv 0$, while its
oscillating part leads to:
\begin{eqnarray}
&&\widetilde{X}_{1}=-\frac{2R_1}{{3}}\widetilde{\xi}_1=
-\frac{2R_1}{{3}}(\widetilde{l}\cos\Phi-\widetilde{\varphi}\sin\Phi)\label{first-order-sol}\\
&&\widetilde{\lambda}_{1}=\frac{2R_0}{{3}}(\widetilde{l}_\tau\cos\Phi-\widetilde{\varphi}_\tau\sin\Phi)\nonumber \\
&&\widetilde{\sigma}_{1}= 2(\widetilde{l}_\tau\sin\Phi+\widetilde{\varphi}_\tau\cos\Phi);\nonumber
\end{eqnarray}
We have used that in dimensionless variables (\ref{scales}),(\ref{scales1})
\begin{eqnarray}\label{RRR}
&&R_0+2R_1=3,\ R_0=3/(1+2\rho),\ R_1=R_2=3\rho/(1+2\rho), \quad \rho\equiv R_1/R_0;
\end{eqnarray}

\noindent \emph{Terms of order} $\varepsilon^2$: Here we consider only the first eqn.(\ref{eqns-tobe-solved1}), which
can be rewritten as
\begin{eqnarray}\label{eqn-Xt}
&&R_0X_t+2R_1x_t=\delta R_1\left\{\frac{2R_0}{l}\left[(1+\xi^2/l^2)(X_t+x_t)+\frac{\xi y}{l^2}y_t\right]
+\frac{R_1x_t}{y}\right\}
\end{eqnarray}
where $x=X+\xi$. Since we consider only linear in $\delta$ precision, then in the second approximation we
should substitute (into the right hand side of (\ref{eqn-Xt})) the solutions from the first equality in
(\ref{first-order-sol}). Further transformations yield
\begin{eqnarray}
&&\overline{X}_{0s}=-\frac{\delta R_1}{{3}}\left[\frac{2R_0(R_0-2R_1)}{{3}}\langle\widetilde{\xi}_1
\widetilde{G}_{1\tau}\rangle+2R_0\langle \widetilde{y}_1 \widetilde{H}_{1\tau}\rangle
+\frac{R_0R_1}{{3}}\langle\widetilde{\xi}_1
\widetilde{K}_{1\tau}\rangle
\right]\label{eqn-Xs}\\
&&G\equiv(1+{\xi^2}/{l^2})/l,\quad H\equiv{\xi y}/{l^3},\quad K\equiv{1}/{y}\nonumber
\end{eqnarray}
It is instructive to write that in the original general notations (\ref{forces-Stokes}) formulae
(\ref{eqn-Xs}),(\ref{RRR}) can be presented as
\begin{eqnarray}\nonumber
&&\overline{X}_{0s}=-\frac{1}{3}\delta\sum_{k=1}^2\sum_{\nu=0}^2
 \sum_{\mu\neq\nu}R_\mu R_\nu \langle \widetilde{S}^{(\nu\mu)}_{1k\tau}\widetilde{x}_{k}^{(\mu)}\rangle
\end{eqnarray}
where both tilde-functions in the right hand side are from the first approximation in $\varepsilon$.

\subsection{Self-propulsion velocity}

The self-propulsion velocity is defined as $\overline{V}_0\equiv\overline{X}_t\simeq\varepsilon^2
\overline{X}_{0s}$. Expressions (\ref{eqn-Xs}),(\ref{RRR}) lead to the formula
\begin{eqnarray}
&&\overline{V}_0=2\varepsilon^2\delta\,U(\Phi,\rho)\langle
\widetilde{l}\widetilde{\phi}_\tau\rangle,\label{V}\\
&& \ U(\Phi,\rho)\equiv\frac{3\rho}{2(1+2\rho)^3}
\left[4\sin\Phi(1-6\rho\cos^2\Phi)+
\frac{\rho}{\sin^2\Phi}\right]\nonumber
\end{eqnarray}
which represents the main result of the paper. Let us discuss it in detail:

(i) $\overline{V}_0$ is proportional to the correlation $\langle
\widetilde{l}\widetilde{\phi}_\tau\rangle$; without any restriction of generality we consider only $\langle
\widetilde{l}\widetilde{\phi}_\tau\rangle>0$.

(ii) Further simplification can be achieved if we accept that the oscillations of arms are harmonic
\begin{eqnarray}\label{harmonic}
\widetilde{l}=\cos(\tau+\theta_1),\quad\widetilde{\varphi}=\cos(\tau+\theta_2);\quad
2\langle\,\widetilde{l}\widetilde{\varphi}_{\tau}\rangle=\sin\theta,\quad \theta\equiv\theta_1-\theta_2
\end{eqnarray}
with constant phases $\theta_1,\theta_2$; $0\leq\theta\leq \pi$. In this case the optimal stroke (providing
the maximum of $\overline{V}_0$) is $\theta=\pi/2$, and
$\max\langle\,\widetilde{l}\widetilde{\varphi}_{\tau}\rangle=1/2$, when the self-propulsion velocity is:
\begin{eqnarray}\label{VV}
\overline{V}_0={\varepsilon^2\delta}\, U(\Phi,\rho)
\end{eqnarray}

(iii) $\overline{V}_0$ (\ref{V}),(\ref{VV}) represents a function of two independent variables $\Phi$ and
$\rho$ for the domain $0<\Phi<\pi$ and $0<\rho<\infty$. Due to symmetry
$\overline{V}_0(\pi/2+\phi,\rho)=\overline{V}_0(\pi/2-\phi,\rho)$ with $0<\phi<\pi/2$, the backward-oriented
(see the figure) $V$-robots with $\Phi=\pi/2- \phi$ and forward-orientated $V$-robots with $\Phi=\pi/2+
\phi$ swim with the same velocity. Hence it is sufficient to study the domain
\begin{eqnarray}\label{domain}
0<\Phi<\pi/2,\quad 0<\rho<\infty,\quad 0<\theta<\pi
\end{eqnarray}
where the restriction $\theta<\pi$ appears due to $\langle \widetilde{l}\widetilde{\phi}_\tau\rangle>0$.

(vi) A sharp singularity $\overline{V}_0\to+\infty$ takes place for $\Phi\to 0$. However this limit does not
have any physical meaning, since small values of $\Phi$ correspond to  collision and overlapping of spheres
$R_1$ and $R_2$. In order to avoid such collision and overlapping, one have to accept
$\sin\Phi>3\delta/2>\delta R_1$. However, a stronger restriction is required in order to provide the validity
of approximation for a velocity field in (\ref{forces-Stokes}), where our basic assumption is: the radius of
each sphere is much smaller than any distance between them; in particular, it means that $\sin\Phi\gg
\delta R_1$. In practice one can take, say,   $\sin\Phi>5\delta$, which for $\delta=0.1$ gives a `secure' domain
instead of  (\ref{domain}):
\begin{eqnarray}\label{domain1}
\pi/6<\Phi<\pi/2,\quad 0<\rho<\infty,\quad 0<\theta<\pi
\end{eqnarray}
More generally, the study of motion with small values of $\Phi$ requires the estimation of errors for the
used approximation; the best way of doing it is computational, see
\cite{Yeomans1}. We do not consider this complex problem and restrict ourselves to
the consideration of internal maxima of $\overline{V}_0$.

(v) $\overline{V}_0\to 0$ when $R_0\to 0$ (or $\rho\to\infty$), which corresponds to the well-known fact that
a dumbbell with oscillating arm is not able to swim; also   $\overline{V}_0\to 0$ when $R_1\to 0$ (or
$\rho\to 0$), which represents a limiting case of one sphere without any oscillations.

(vi) A local maximum of $\overline{V}_0$ for any $\rho=\const$ always takes place at $\Phi=\pi/2$, when
\begin{eqnarray}\label{V1}
U(\pi/2,\rho)={3\rho(4+{\rho})}/{[2(1+2\rho)^3]}
\end{eqnarray}
Hence, a completely `open' $V$-robot (with  $2\Phi=\pi$) swims in positive direction (see the figure) with
the maximal speed. Function $U(\pi/2,\rho)$ (\ref{V1}) is increasing linearly for small $\rho$; $U$ reaches
$\max U\simeq 0.47$ at $\rho\simeq 0.27$, and then $U$ decreases to zero rapidly and monotonically. Hence we
can write that
\begin{eqnarray}\label{Vmax}
\max \overline{V}_0(\Phi,\rho,\theta)\simeq 0.47\epsilon^2\delta\quad \text{at}\quad\Phi=\pi/2,\quad
\rho\simeq 0.27,\quad
\theta=\pi/2
\end{eqnarray}
It is interesting to compare this result with the result  for a homogeneous linear three-sphere
micro-swimmer, where $\max\overline{V}_0\simeq 0.19\epsilon^2\delta$ (see \cite{VladimirovX3}). The
comparison shows that $V$-robot swims about 2.5 times faster than a linear micro-robot (when in both cases
the strokes are harmonic and optimal). The fastest swimming of $V$-robot takes place when the spheres $R_1$
and $R_2$ are approximately $4$ times smaller than $R_0$.

(vii) $V$-robot can swim in negative direction (see the figure) at $\Phi=\pi/4$, when the function
$U(\pi/4,\rho)$ is non-monotonic and changes its sign: $U$ increases linearly for small $\rho$, then $U$
reaches $\max U\simeq 0.19$ at $\rho\simeq 0.13$, after that decreases such that $U=0$ at $\rho=0.43$. For
$\rho>0.43$ we have $U<0$ and  $\min U=U(1.75)\simeq -0.24$ which yields
\begin{eqnarray}\label{Vmin}
\min \overline{V}_0(\pi/4,\rho,\pi/2)\simeq -0.24\, \epsilon^2\delta\quad \text{at}\quad
\rho\simeq 2.10
\end{eqnarray}
It follows from (\ref{V}) that $\min \overline{V}_0$ in all domain (\ref{domain}) is
\begin{eqnarray}\label{VminA}
\min \overline{V}_0(\Phi, \rho,\theta)\simeq -0.26\, \epsilon^2\delta\quad \text{at}\quad
\Phi=0.69,\quad \rho\simeq 1.65,\quad \theta=\pi/2
\end{eqnarray}
which is close to (\ref{Vmin}). Hence $V$-robot swims in negative direction if the angle between arms is
close to $\pi/2$ (or $\Phi=\pi/4$). The maximal speed of this reverse swimming is approximately two times
slower than the maximal speed in positive direction.

(viii) It is also  of interest that for three equal spheres ($\rho=1$) we have
\begin{eqnarray}\label{V1}
&&\max \overline{V}_0(\Phi,1,\theta)\simeq 0.28\epsilon^2\delta\quad \text{at}\quad\Phi=\pi/2,\quad
\theta=\pi/2\nonumber\\
&&\min \overline{V}_0(\Phi,1,\theta)\simeq -0.23\, \epsilon^2\delta\quad \text{at}\quad
\Phi=0.68,\quad \theta=\pi/2\nonumber
\end{eqnarray}
which shows an essential reduction of the speed in positive direction.

(ix) The calculations at the boarder $\Phi=\pi/6$ of a `secure' domain (\ref{domain1}) show that
$U(\pi/6,\rho)$ increases linearly for small $\rho$, reaches  $\max U\simeq 0.13$ at $\rho\simeq 0.13$, then
decreases such that $U(0.40)\simeq 0$ and $U<0$ for $\rho>0.40$;  $\min U=U(1.65)\simeq -0.18$. These values
show that $\Phi=\pi/6$ is still well away of a singularity at $\Phi\to 0$.

\subsection{Power and efficiency}

The power of V-robot is defined as
\begin{eqnarray}\label{P1}
\mathcal{P}\equiv\sum_{\nu=0}^2\vf^{(\nu)}\cdot\vx_t^{(\nu)}
\end{eqnarray}
where $\vf^{(\nu)}$ is the force, exerted by the arms on the $\nu$-th sphere. The total force exerted on each
sphere must be zero, hence $\vf^{(\nu)}+\vF^{(\nu)}=0$ with the friction force $\vF^{(\nu)}$
(\ref{forces-Stokes}). Therefore the  main term of (\ref{P1}) can be presented as
\begin{eqnarray}\label{P2}
\mathcal{P}\simeq\sum_{\nu=0}^2R_\nu\,\vx_t^{(\nu)}\cdot\vx_t^{(\nu)}
\simeq\varepsilon^2\sum_{\nu=0}^2R_\nu\,\widetilde{\vx}_{1\tau}^{(\nu)}\cdot\widetilde{\vx}_{1\tau}^{(\nu)}
\end{eqnarray}
where we have used (\ref{x-f-ser1}). Then the use of (\ref{first-order}),(\ref{notat-1}),(\ref{oper-1})
yields
\begin{eqnarray}\label{P3}
\overline{\mathcal{P}}\simeq\frac{3\varepsilon^2}{(1+2\rho)}\langle\widetilde{X}_{1\tau}^2+2\rho(\widetilde{x}_{1\tau}^2+
\widetilde{y}_{1\tau}^2)\rangle
\end{eqnarray}
and the use of (\ref{first-order-sol}) leads to
\begin{eqnarray}\label{P4}
\overline{\mathcal{P}}\simeq\frac{6\varepsilon^2\rho}{(1+2\rho)^2}\langle\widetilde{l}_{\tau}^2+\widetilde{\varphi}_{\tau}^2+
2\rho[\widetilde{l}_{\tau}^2\sin^2\Phi+\widetilde{\varphi}_{\tau}^2\cos^2\Phi+
\widetilde{l}_{\tau}\widetilde{\varphi}_{\tau}\sin(2\Phi)]\rangle
\end{eqnarray}
For harmonic oscillations (\ref{harmonic}) it gives
\begin{eqnarray}\label{P5}
\overline{\mathcal{P}}\simeq\frac{6\varepsilon^2\rho}{(1+2\rho)^2}W,\quad W\equiv[1+
\rho\left(1+\sin(2\Phi)\cos\theta\right)]
\end{eqnarray}
Another expression $\overline{\mathcal{P}}_s=3\overline{V}_0^2$  represents the power, which is required to
drag  $V$-robot with velocity $\overline{V}_0$ in the absence of its oscillations when the main approximation
for the dimensionless Stokes friction force is $-3\overline{V}_0$ (where the coefficient 3 represents the sum
of all radii (\ref{RRR})). Lighthill's swimming efficiency (see \cite{Koelher}) is the ratio
$\mathcal{E}\equiv{\overline{\mathcal{P}}_s}/{\overline{\mathcal{P}}}$. For harmonic oscillations
(\ref{harmonic}) expressions (\ref{VV}),(\ref{P5}),(\ref{V}) give
\begin{eqnarray}\label{P6}
\mathcal{E}=\mathcal{E}(\Phi,\rho,\theta)\simeq  \varepsilon^2\delta^2 \frac{9\rho}{8(1+2\rho)^4} \frac{U^2}{W}\sin^2\theta
\end{eqnarray}
The analysis and the computations show that
\begin{eqnarray}\label{Emax}
\max\mathcal{E}\simeq 0.89   \varepsilon^2\delta^2,\quad\text{at} \quad \Phi= \pi/2,\quad
\rho\simeq 0.15,\quad \theta= \pi/2
\end{eqnarray}
One can see that $\max\mathcal{E}$ takes place at the same $\Phi= \pi/2$ and $\theta= \pi/2$, but for
approximately twice smaller $\rho$ than for $\max \overline{V}_0$ (\ref{Vmax}); hence the most efficient
swimming takes place when the spheres $R_1$ and $R_2$ are approximately $7$ times smaller than $R_0$. It is
also interesting that for a linear three-sphere micro-robot $\max\mathcal{E}\simeq 0.18
\varepsilon^2\delta^2$ (see
\cite{VladimirovX3}), which is five times lower than (\ref{Emax}).

\section{Discussion}

(i) V-robot has already been  studied numerically by \cite{Yeomans1}, but never analytically. Quantitative
comparison between our analytical results and computations of
\cite{Yeomans1} is difficult, since these authors studied the large amplitudes of arms' oscillations and
non-harmonic strokes; also the domain of the main parameters, they studied, is unclear from the text. At the
same time the existence of both forward (\ref{Vmax}) and reverse (\ref{VminA}) swimming  provides qualitative
agreement of our results with
\cite{Yeomans1}.

(ii) One can see that the magnitude of velocity in terms of small parameters
$\overline{V}_0=O(\varepsilon^2\delta)$ (\ref{V}) is the same as the result by
\cite{Golestanian, Golestanian1} for a linear three-sphere robot and by \cite{VladimirovX3} for a linear $N$-sphere
robot. At the same time our choice of slow time $s=\varepsilon^2 t$ (\ref{chain}) agrees with classical
studies of self-propulsion for low Reynolds numbers, see \cite{Taylor, Blake, Childress}, as well as the
geometric studies of
\cite{Wilczek}.

(iii) In our examples, all arms move harmonically (\ref{harmonic}); it does not provide the maximum of
$\overline{V}_0\sim\langle\,\widetilde{l}\widetilde{\varphi}_{\tau}\rangle$ (\ref{V}), see the relevant
discussion in \cite{VladimirovX3}.The studies of non-periodic oscillations represent an interesting
additional problem, see \cite{Golestanian1}.

(iv) In our study we build an asymptotic procedure with two small parameters: $\varepsilon\to 0$ and
$\delta\to 0$. Such a setting usually requires the consideration of different asymptotic paths on the plane
$(\varepsilon,\delta)$ when, say $\delta=\delta(\varepsilon)$. In our case we can avoid such  consideration,
since  small parameters appear (in the main order) as a product $\varepsilon^2\delta$.

(v) The mathematical justification of the presented results by the estimation of an error in the original
equation can be performed similar to \cite{VladimirovX1,VladimirovX2}. It is also possible to derive the
higher approximations of $\overline{V}_0$, as it has been done by
\cite{VladimirovX1,VladimirovX2} for different cases. The higher approximations can be useful for the studies
of motion with $\overline{V}_0\equiv 0$ (which are always possible for $V$-robots).

(vi) In order to compare the velocities of micro-robots and micro-organisms we use the dimensional variables,
in which $\max\overline{V}_0^*\simeq 0.47 \omega^* L^*
\varepsilon^2\delta$; it  shows that $V$-robot can move itself with the speed  about $10$ percent
of its own size per second (we have taken $\varepsilon=\delta=0.2$ and $\omega^*=30 s^{-1}$; the value of
$\omega^*$ can be found in \cite{PedKes, VladPedl, Pedley, Polin}. From these papers we also can see that
this estimation of $\overline{V}_0^*$ is about $10$ times lower than a similar value for natural
micro-swimmers.

\begin{acknowledgments}
The author is grateful to Profs. M. Bees,  K.I. Ilin, H.K.Moffatt, T.J.Pedley, and J. Pitchford for useful
discussions.
\end{acknowledgments}


\begin{thebibliography}

\bibitem[Alexander \emph{et.al.}  (2009)]{Yeomans} \textsc{Alexander, G. P., Pooley, C. M., and Yeomans, J.M.}
2009 Hydrodynamics of linked sphere model swimmers. {\it J. Phys.: Condens. Matter}, \textbf{21}, 204108.

\bibitem[Alouges \emph{et.al.} (2008)]{Lefebvre} \textsc{Alouges, F., DeSimone, A., and Lefebvre, A.}
2008 Optimal strokes for low Reynolds number swimmers: an example. {\it J. Nonlinear Sci.},
\textbf{18}, 277-302.

\bibitem[Avron \emph{et.al.} (2005)]{Avron} \textsc{Avron, J.E., Kenneth, O., and Oaknin, D.H.}
2005 Pushmepullyou: an efficient micro-swimmer. {\it New J. of Physics},
\textbf{7}, 234.

\bibitem[Becker \emph{et.al.}  (2003)]{Koelher}
\textsc{Becker, D. J., Koelher, C. M., Ryder, and Stone, J.M.}
2003 On self-propulsion of micro-machimes al low Reynolds number: Purcell's three-link swimmer. {\it J. Fluid
Mech.}, \textbf{490}, 15-35.

\bibitem[Blake  (1971)]{Blake}
\textsc{Blake, J. R.} 1971 Infinite models for ciliary propulsion. {\it J. Fluid Mech.},
\textbf{49}, 209-227.

\bibitem[Chang \emph{et.al.} (2007)]{Paunov}
\textsc{Chang, S.T., Paunov, V.N., Petsev, D.N., and Orlin, D.V.}
2007 Remotely powered self-propelling particles and micropumps based on miniature diodes. {\it Nature
Materials}, \textbf{6}, 235-240.

\bibitem[Childress (1981)]{Childress} \textsc{Childress, S.} 1981 \emph{Mechanics of swimming and flying.}
Cambridge, CUP.


\bibitem[Dreyfus \emph{et.al.} (2005)]{Dreyfus}
\textsc{Dreyfus , R., Baudry, J., Roper, M.L., Fermigier, M., Stone, H.A. and Bibette, J.}
2005 Microscopic artificial swimmers. {\it Nature},
\textbf{437}, 6, 862-865.

\bibitem[Earl \emph{et.al.} (2007)]{Yeomans1}
\textsc{Earl, D. J., Pooley, C. M., Ryder, J.F., Bredberg, I. and Yeomans, J.M.}
2007 Modelling microscopic swimmers at low Reynolds number. {\it J. Chem. Phys.},
\textbf{126}, 064703.


\bibitem[Gilbert \emph{at.al.} (2010)]{Gilbert}
\textsc{Gilbert, A. D., Ogrin, F. Y., Petrov, P.G., and Wimlove, C.P.}
2010 Theory of ferromagnetic microswimmers. {\it Q.Jl Mech. Appl. Math.},
\textbf{64}, 3, 239-263.

\bibitem[Golestanian \& Ajdari (2008)]{Golestanian} \textsc{Golestanian, R. and Ajdari, A.} 2008
Analytic results for the three-sphere swimmer at low Reynolds number. {\it Phys.Rev.E}, \textbf{77}, 036308.

\bibitem[Golestanian \& Ajdari (2009)]{Golestanian1} \textsc{Golestanian, R. and Ajdari, A.} 2009
Stochastic low Reynolds number swimmers. {\it J. Phys.: Condens. Matter},
\textbf{21}, 204104.


\bibitem[Landau \& Lifshitz (1959)]{Landau} \textsc{Landau, L.D. and Lifshitz, E.M.} 1959
\emph{Fluid Mechanics.} Oxford, Butterworth-Heinemann.

\bibitem[Lamb (1932)]{Lamb} \textsc{Lamb, H.} 1932 \emph{Hydrodynamics.} Sixth edition,
Cambridge, CUP.


\bibitem[Lauga (2011)]{Lauga}  \textsc{Lauga, E.} 2011 Life around the scallop theorem. {\it Soft Matter},
\textbf{7}, 3060-3065.


\bibitem[Leoni \emph{et.al.} (2009)]{Pietro}
\textsc{Leoni, M., Kotar, J., Bassetti, B., Cicuta, P. and Lagomarsino, M.C.}
2009 A basic swimmer at low Reynolds number. {\it Soft Matter},
\textbf{5}, 472-476.

\bibitem[Moffatt (1996)]{Moffatt} \textsc{Moffatt, H. K.} 1996 \emph{Dynamique des Fluides, Tome 1,
Microhydrodynamics.} Ecole Polytechnique, Palaiseau.

\bibitem[Najafi \& Golestanian (2004)]{NG+} \textsc{Najafi, A. and Golestanian, R.} 2004
Simple swimmer at low Reynolds number: three linked spheres. {\it Phys.Rev.E}, \textbf{69}, 062901.

\bibitem[Pedley \& Kessler (1987)]{PedKes} \textsc{Pedley, T.J. and Kessler, J.O.} 1987
The orientation of spheroidal microorganisms swimming in a flow field. {\it Proc. R. Soc. Lond.},
B\textbf{231}, 47-70.

\bibitem[Pedley (2009)]{Pedley}  \textsc{Pedley, T.J.} 2009 Biomechanics of aquatic micro-organisms.
{\it New trends in fluid mechanics research}, \textbf{1}, 1-6.

\bibitem[Polin \emph{et.al.} (2009)]{Polin}
\textsc{Polin, M., Tuval, I., Dresher, K., Gollub, J. P. and Goldstein, R.E.}
2009 \emph{Chlamydomonas} swims with two gears in a eukaryotic version of run-and-tumble locomotion. {\it
Science}, \textbf{325}, 487-490.

\bibitem[Purcell (1977)]{Purcell} \textsc{Purcell, E.M.} 1977
Life at low Reynolds number. {\it Amer. J. of Phys.}, \textbf{45}, 1, 3-11.


\bibitem[Shapere \& Wilczek (1989)]{Wilczek} \textsc{Shapere, A. and Wilczek, F.} 1989
Efficiencies of self-propulsion at low Reynolds number. {\it J. Fluid Mech.},
\textbf{198}, 587-599.

\bibitem[Taylor (1951)]{Taylor} \textsc{Taylor, G.
I.} 1951, Analysis of the swimming of microscopic organisms. {\it Proc. R. Soc. Lond.}, \textbf{A209},
447-461.

\bibitem[Vladimirov \emph{et.al.} (2004)]{VladPedl}
\textsc{Vladimirov, V.A., Wu, M.S., Pedley, T.J., Denissenko, P.V. and Zakhidova S.G.}
2004 Measurements of cell velocity distributions in populations of motile algae. {\it J. Exp. Biol.},
\textbf{207}, 1203-1216.

\bibitem[Vladimirov (2005)]{Vladimirov0} \textsc{Vladimirov, V.A.}
2005  Vibrodynamics of pendulum and submerged solid.  {\it J. of Math. Fluid Mech.} \textbf{7}, S397-412.

\bibitem[Vladimirov (2008)]{Vladimirov1} \textsc{Vladimirov, V.A.}
2008 Viscous flows in a half-space caused by tangential vibrations on its boundary. {\it Studies in Appl.
Math.}, \textbf{121}, 4, 337-367.



\bibitem[Vladimirov (2010)]{VladimirovX1} \textsc{Vladimirov, V.A.} 2010
Admixture and drift in oscillating fluid flows. E-print:
\emph{ArXiv}: 1009.4085v1, (physics,flu-dyn).


\bibitem[Vladimirov (2011)]{VladimirovX2} \textsc{Vladimirov, V.A.} 2011
Theory of non-degenerate oscillatory flows. E-print: \emph{ArXiv}: 1110.3633v2, (physics,flu-dyn).

\bibitem[Vladimirov (2012a)]{VladimirovMHD} \textsc{Vladimirov, V.A.}
2012a  Magnetohydrodynamic drift equations: from Langmuir circulations to magnetohydrodynamic dynamo?  {\it
J. Fluid Mech.} \textbf{698}, 51-61.

\bibitem[Vladimirov (2012b)]{VladimirovX3} \textsc{Vladimirov, V.A.} 2012b
On self-propulsion of $N$-sphere micro-robot. E-print: \emph{ArXiv}: 1206.0890v1 and 1209.0171v1,
(physics,flu-dyn); also submitted to the \emph{Journal of Fluid Mechanics}.

\end{thebibliography}
\end{document}